# Towards Recommender Systems for Police Photo Lineup

Ladislav Peska
Department of Software Engineering
Faculty of Mathematics and Physics, Charles University, Prague
Czech Republic
peska@ksi.mff.cuni.cz

Hana Trojanova
Department of Psychology
Faculty of Arts, Charles University, Prague
Czech Republic
trojhanka@gmail.com

## ABSTRACT

Photo lineups play a significant role in the eyewitness identification process. This method is used to provide evidence in the prosecution and subsequent conviction of suspects. Unfortunately, there are many cases where lineups have led to the conviction of an innocent suspect. One of the key factors affecting the incorrect identification of a suspect is the lack of lineup fairness, i.e. that the suspect differs significantly from all other candidates. Although the process of assembling fair lineup is both highly important and time-consuming, only a handful of tools are available to simplify the task.

In this paper, we describe our work towards using recommender systems for the photo lineup assembling task. We propose and evaluate two complementary methods for item-based recommendation: one based on the visual descriptors of the deep neural network, the other based on the content-based attributes of persons.

The initial evaluation made by forensic technicians shows that although results favored visual descriptors over attribute-based similarity, both approaches are functional and highly diverse in terms of recommended objects. Thus, future work should involve incorporating both approaches in a single prediction method, preference learning based on the feedback from forensic technicians and recommendation of assembled lineups instead of single candidates.

## 1 INTRODUCTION

Evidence from eyewitnesses often plays a significant role in criminal proceedings. A very important part is the lineup - eyewitness identification of the perpetrator (either lineup in natura, or a photo lineup, i.e., demonstration of objects' photographs). A lineup is assembled by placing a suspect within a group of known innocent persons (fillers). Lineups may lead to the prosecution and subsequent conviction of the perpetrator. Yet there are cases where lineups can play a role in the conviction of an innocent suspect [20].

This forensic method consists of the recognition of persons or things and thus is linked with a wide range of psychological processes such as perception, memory, and decision making. Those processes can be influenced by the lineup itself. In order to prevent witnesses from making incorrect identifications, the lineup assembling task is for several decades among the top research topics of the psychology of eyewitness identification [2, 3, 4, 5, 14, 15, 19, 20, 21]. The sources of error in eyewitness identifications are numerous. Some variables are based on the recalled past event (e.g., distance from the scene, lighting conditions) and context of the witness (e.g., level of attention, age or different ethnicity of the witness and the suspect). Such features cannot be controlled in general [5, 15, 19]. Controllable variables include the method of questioning, construction of the lineup, interaction with investigators, and more [5, 19, 21].

One of the principal recommendations for inhibiting errors in identification is to assemble lineups according to the lineup fairness principle [2, 14]. Roughly speaking, fair lineups should ensure that the suspect is not substantially different from the fillers [23]. All fillers should have similar visual attributes to the suspect to provide an appropriate level of uncertainty to unreliable eyewitnesses[1]. Eyewitnesses are more confident in their identification when fillers are too different, even in cases where incorrect suspect is presented [3]. Since the similarity metric is latent, lineup fairness is usually assessed on the basis of data obtained from "mock witnesses", i.e., persons who have not seen the offender, but received a short description of him/her. Lineup is considered fair (also denoted as unbiased) if mock witnesses are unable to identify a suspect based only on a brief textual description. See Fig. 1 for an example of a highly biased lineup.

Assembling fair lineups, i.e., selecting suitable lineup fillers for a particular suspect is a challenging and time-consuming task involving the exploration of large datasets of persons. In the recent years, some research projects [13, 24] as well as commerce activities, e.g., *elineup.org*, aimed to simplify the process of eyewitness identifications. However, they mostly focused on the lineup administration and the support of lineup assembling is at best at the level of attribute-based searching of candidates' database.

From the point of view of recommender systems, lineup assembling is quite specific task for several reasons. Users of the system are respected experts, who assemble lineups regularly, although, usually, not on a daily bases. Thus, we can expect a steady flow of feedback from long-term users. Also, each lineup assembling task is highly unique, i.e., the same suspect hardly ever appears in multiple lineups. Thus, the approaches based solely on collaborative filtering [6] cannot be applied in this scenario. Last, but not least, the relevance judgement is highly dependent on the visual appearance and/or visual similarity of the suspect and candidates. This observation the chance to deploy some visual similarity learning methods, e.g., deep neural networks for this task. In recent years, deep neural networks (DNN) were successfully applied on tasks ranging from general image classification [12] and multimedia retrieval [16] to session-based recommendations [22] and face recognition problems [7, 9, 17] and clearly defined new state-of-the-art on multimedia retrieval. However, previous approaches mostly focused on identification or categorization of persons and objects and it was not yet shown, that they can also induce rather than modelling relevant inter-persons similarity.

---

[1] https://psychology.iresearchnet.com/forensic-psychology/eyewitness-memory/lineup-size-and-bias/

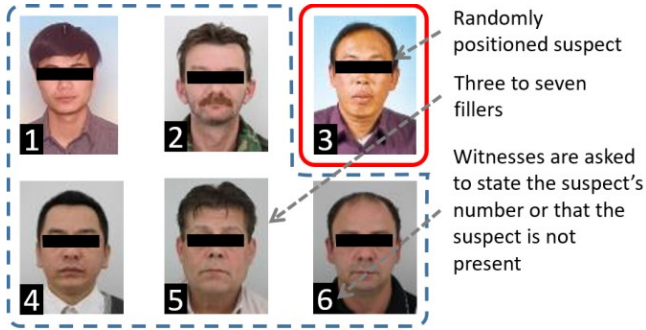

**Figure 1:** Example of an extremely biased lineup. Lineup usually consists of four to eight persons and witness is instructed that suspect may or may not be among them. However in this case, suspect can be easily identified even by a mock witness knowing only a short description such as, "Vietnamese male, 50-70 years old."

**Table 1:** Lineup dataset characteristics. The dataset contains a total of 4,423 persons; the table entries depict the frequency of selected features.

| Top nationalities (84 in total) | | | |
|---|---|---|---|
| Czech | Vietnamese | Ukrainian | Slovak |
| 63.0% | 8.3% | 6.7% | 5.5% |
| **Age groups** (dataset also contains exact age) | | | |
| 0-18 | 18-35 | 35-55 | 55+ |
| 2.1% | 27.1% | 51.3% | 11.0% |
| **Top appearance features** (441 in total) | | | |
| Average figure | 52,7% | Black hair | 19,2% |
| Brown eyes | 39,3% | Blue eyes | 16,2% |
| Straight hair | 34,5% | Curly hair | 15,0% |
| Thin figure | 32,2% | Green eyes | 6,7% |
| Black-brown hair | 30,4% | Blond hair | 4,9% |

[2] e.g., *http://elineup.org, http://www.crimestar.com*
[3] *http://aplikace.policie.cz/patrani-osoby/Vyhledavani.aspx*

In this paper, we describe our work in progress towards designing recommender systems for the fair lineup assembling problem. As this is an initial work on this domain, we focused on utilizing existing techniques to this novel application scenario, evaluate them and describe its potential extensions and improvements. We propose two item-based methods to recommend lineup candidates for a particular suspect. Both proposed methods were evaluated by domain experts with respect to the *intrinsic fairness metric* in a realistic user-study. The main contributions of this paper are:

- Proposed methods for item-based recommendations of candidates for lineup assembling task.
- Evaluation of both methods in lineup assembling user-study by forensic technicians.
- Published dataset of candidates and assembled lineups available for future work.

To the best of our knowledge, this is the first approach on applying principles of recommender systems on lineup assembling problems with respect to the lineup fairness metric.

## 2 MATERIALS AND METHODS

### 2.1 Dataset of Lineup Candidates

A necessary precondition for assembling fair lineups is a suitable database of lineup candidates. Although there are some commercial lineup databases[2], we need to approach carefully while applying such datasets due to the problem of localization. Not only are the racial groups highly different e.g., in North America (where the datasets are mostly based) and Central Europe, but other aspects such as common clothing patterns, haircuts or make-up trends vary greatly in different countries and continents. Thus, underlined datasets should follow the same localization as the suspect in order to inhibit the bias of detecting strangers or having the incorrect ethnicity in the lineup.

We evaluated the proposed methods in the context of the Czech Republic. Although the majority of the population is Caucasian, mostly of Czech, Slovak, Polish and German nationality, there are large Vietnamese and Romany minorities which makes lineup assembling more challenging. Another reason for Czech localization was that we were able to collect a sufficiently large database of lineup candidates through the *wanted and missing persons* application[3] of the Police of the Czech Republic. In total, we collected data about 4,423 missing or wanted males. All records contained a photo, nationality, age and appearance characteristics such as: (facial) hair color and style, eye color, figure shape, tattoo, scars and more. Some dataset statistics are depicted in Table 1.

### 2.2 Recommending Strategies for Lineup Assembling

Both proposed recommending strategies are based on the assumption that lineup fairness can be approximated through the similarity of the suspect and fillers, i.e. by filling lineups with candidates similar to the suspect, we ensure that lineups remain unbiased[4]. Proposed methods differ in the underlined similarity model. In the first case, the assumption is that similarity of candidates can be approximated by the similarity of their attributes. The later method supposes that candidates' similarity can be approximated through the visual descriptors based on their photography.

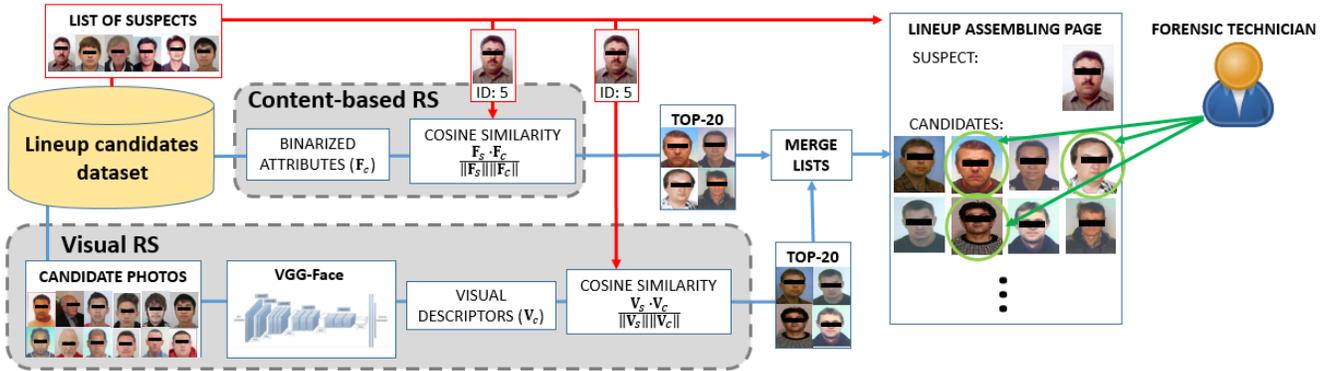

**Figure 2:** Outline of the evaluation protocol. For each suspect, both methods recommend top-20 lineup candidates based on the internal similarity metric. Both lists are merged and shown to the forensic technician, who selects most suitable fillers.

*2.2.1 Content-based Recommendation Strategy (CB-RS).* The first proposed strategy leverages the collected content-based attributes of candidates. We employed the Vector Space Model [10] with binarized content-based features, TF-IDF feature weighting and candidate similarity defined as the cosine similarity of candidates' feature vectors. *CB-RS* strategy is intended to be closely similar to the attribute-based searching, which is commonly available for the lineup assembling task. Thus, *CB-RS* can be considered as a baseline method.

*2.2.2 Recommendation Based on Deep Neural Network (Visual-RS).* The second proposed method leverages the similarity of visual descriptors received from a pre-trained DNN. More specifically, we utilized a VGG network for facial recognition problems (VGG-Face, [17]). The authors approached the training as a multi-classification problem on the dataset of 2,622 celebrities with 1,000 images each. This method achieved state-of-the-art performance ratings on several face recognition benchmarks. We used the *probability layer* of the VGG-Face network as a visual descriptor for each lineup candidate and defined the similarity of candidates as a cosine similarity of their visual descriptors.

## 3 EVALUATION AND RESULTS

During the evaluation phase, we wanted to clarify several research questions. First of all, as this is an initial research on applying recommender systems for lineup assembling, it is necessary to illustrate the applicability of the approach. The acceptance question is whether the problem of assembling unbiased lineups can be approached from the perspective of recommender systems through the similarity of candidates with a suspect. If so, we intend to evaluate the results of both recommendation strategies and whether their performance differs for some particular groups of suspects.

The ultimate goal of recommender systems in lineup assembling is to move from recommending candidates towards automated recommendation of assembled lineups. A necessary step towards this task is the acceptance of the candidates ordering by the lineup administrator, i.e., whether the latent similarity metric applied by the administrator can be modeled by one of the recommending strategies.

In the early stages of the project, we will approximate the actual lineup fairness (evaluated by mock witnesses) through the user study on lineup assembling performed by domain experts, i.e., forensic technicians.

### 3.1 Evaluation Protocol

For the purpose of the user study, we randomly selected 30 persons from the lineup candidates' dataset to perform the role of suspects. For each suspect, both recommendation strategies proposed top-20 candidates. Surprisingly, the candidate sets of both methods were almost completely different with the average intersection size of 1.83%. The proposed candidates from both methods were randomly merged into a single list[5] and displayed together with the suspect to the domain experts.

The task of the user study was as follows: for a given suspect, select suitable fillers from the list of candidates so that the resulting lineup is unbiased. Participants were instructed to maintain lineup fairness principles, they were allowed to produce incomplete lineups if no more suitable candidates were available, or select more candidates if they were equally eligible. Participants did not know which recommendation strategy proposed which candidate. The evaluation protocol is illustrated on Fig. 2. The evaluation task was followed by a questionnaire and moderated discussion on whether the participants were able to select (a sufficient amount of) relevant candidates from the proposed ones and if they could point out some obvious errors in the recommendations. The participants were also asked whether they would prefer more diverse lists of candidates, would like to view further candidates (another page), would like to be able to view similar persons to a particular candidate, or would rather be able to use an attribute search.

---

[4] In this stage of research we do not consider the problem of too similar objects, e.g., twins of a suspect as those are rather rare in the underlined dataset. However, eliminating near-duplicates strategy can be used to alleviate this problem.

[5] The ordering of candidates proposed by each method was maintained, i.e., the randomness was applied on the decision, whether the next list item will be filled by *CB-RS* or *Visual-RS* method.

Due to the low number of participants, the questionnaire did not provide statistically significant results by itself, however, it supported claims based on the evaluation results in several cases and revealed some surprising observations mentioned in the discussion.

## 3.2 Results and Discussion

The evaluation was performed by seven forensic technicians from the Czech Republic, with 202 assembled lineups in total. Participants selected in total 800 fillers, out of which 298 were proposed by content-based recommending strategy, 466 were proposed by visual recommending strategy and 36 selected persons were proposed by both recommending strategies. The average number of selected persons was 4.0 ($\sigma = 0.9$). Based on this and on the follow-up questionnaire, we can conclude that using recommender systems in the lineup assembling task is a viable strategy. However, there is room for improvement as lineups should optimally contain five fillers [18] accompanying the suspect and the participants were rarely able to select the optimal volume of fillers. This was also reflected in the questionnaire, where participants mentioned that there were relevant candidates, but sometimes not enough to assemble the whole lineup.

As for particular evaluation strategies, *Visual-RS* clearly outperformed *CB-RS* in general and when considering only suspects from Central Europe. This is also supported by the higher level of agreement on selected candidates of the *Visual-RS* method. The performance difference, however, was much smaller while considering suspects from outside of Central Europe (Vietnamese, Azerbaijani, Kyrgyz, Ukrainian and Serbian nationalities). See Table 2 for more details.

We hypothesize that such behavior can be attributed to two aspects. First, several researchers pointed out the effects of another race in face recognition [1, 15]. In the described experiments, participants of one ethnicity were less able to distinguish or identify persons from different ethnic group than from their own. Such effect could emerge also in cases of lineup administrators (all of them were from Central Europe), who could compare general characteristics (e.g., ethnicity or age group) of the suspect and candidates, but did not recognize other distinctive features of the particular ethnic[6]. In such cases, *CB-RS* fits well with the participants' capability as nationality and age groups are incorporated within the background knowledge. This is also illustrated by increased level of agreement for *CB-RS*. Furthermore, the VGG-Face network was not trained to distinguish between different nationalities or age groups, and thus it recommended candidates with similar facial features who were of a different nationality or age. This problem was also noted during the follow-up questionnaire, so future work should incorporate either fine-tuning (several variants) of VGG-Face network to distinguish between some content-based attributes [9], or employ attribute-based similarity as pre/post-filtering.

---

[6] While evaluating face, white participants are fixated mainly on the eye region and e.g. Asian participants are fixated more on the central region of the face [1].

**Table 2:** Evaluation results depicting the volume of lineups, selected candidates, the differences in volumes of selected candidates per lineup (p-value of paired t-test) and the level of agreement on selected candidates among user-study participants (Krippendorff's alpha).

|  | Total lineups | Selected candidates | Paired t-test | Level of agreement |
|---|---|---|---|---|
| **All lineups** | | | | |
| Visual-RS | 202 | **466** (58%) | 1.2e-8 | 0.178 |
| CB-RS | | 298 (37%) | | 0.138 |
| **Lineups with suspects from Central Europe** | | | | |
| Visual-RS | 149 | **361** (61%) | 2.1e-8 | 0.197 |
| CB-RS | | 216 (36%) | | 0.113 |
| **Lineups with suspects outside Central Europe** | | | | |
| Visual-RS | 53 | 105 (51%) | 0.104 | 0.128 |
| CB-RS | | 82 (40%) | | 0.205 |

The mean rank of selected candidates was 8.2 ($\sigma = 5.7$) in case of *Visual-RS* and 8.9 ($\sigma = 5.2$) for *CB-RS*, so both methods seemingly provided relevant ordering also for the top-20 candidates. However, the effect of search engine bias [8] could also contribute to the perceived values and should be evaluated in a separate experiment.

The participants' questionnaire revealed another interesting aspect of photo lineup assembling problems. Whereas the diversity of the recommended list is of high importance in many recommender systems application, participants of the evaluation did not require more diverse recommendations. Furthermore, several participants mentioned the need for uniformity in the final lineup, i.e. that the intra-list diversity is as small as possible. An iterative recommender system can be built upon this observation, where the list of recommended candidates will be gradually refined to be similar both with the suspect and already selected candidates.

Finally, the candidate selection can be considered as an implicit feedback and leveraged to model either individual or global preferences. The need for individual preferences can be observed from the relatively high level of disagreement among participants on the selected candidates for each lineup (see Table 2). Although the domain fragmentation will prevent us from employing collaborative filtering approaches, the expected long-term usage combined with rich data received from multiple sources enables content-based preference learning models. However, we need to approach personalization carefully as the individual administrator's biases (i.e. systematic deviations from fair lineups) may be exacerbated by the personalization technique.

Implicit feedback can be used to fine-tune DNN or in some feature-weighting approach [11] to distinguish selected and ignored candidates on the level of content-based descriptors.

## 4 CONCLUSIONS

The main aim of this work in progress is to analyze the applicability of recommender systems principles in the problem of photo lineup assembling. Although the photo lineup assembling task is time-consuming and of large importance, the state-of-the-art methods so far provided merely attribute search API and to the best of our knowledge, no recommending approach was ever applied to the lineup assembling task.

As an initial approach, we proposed two item-based recommending strategies: *CB-RS* and *Visual-RS*. The evaluation based on domain experts showed that *Visual-RS* significantly outperformed *CB-RS* recommendations and that recommender systems in general provides substantial aid to the user during the lineup assembling.

Both methods proposed a substantial volume of relevant candidates and some of the specific errors produced by each method can be detected by the other, so a merged recommendation strategy should be incorporated in future work. The evaluation also pointed out some specifics of the photo lineup domain, such as: the diverse effect of foreign ethnicities on recommending strategies, the need for lineup intra-list similarity and the importance of modeling the individual preferences of the lineup administrators. Thus, a substantial part of our future work should focus on incorporating these observations into improved recommending strategies and evaluating fairness of proposed lineups directly via mock witnesses.

## ACKNOWLEDGMENTS

This work was supported by the Czech grants GAUK-232217, GACR-17-22224S and P46. Some additional materials are available online:

- Source codes of the dataset collection script and recommendation methods: *http://github.com/lpeska/lineups*

- Dataset of lineup candidates, evaluated lineups and selected persons: *https://tinyurl.com/yat3rtr2*